\documentclass[journal,compsoc]{IEEEtran}
\usepackage{flushend}

\ifCLASSOPTIONcompsoc
  \usepackage[nocompress]{cite}
 
\else
  \usepackage{cite}
\fi
 \usepackage{graphicx}

\ifCLASSINFOpdf
\else
\fi
\usepackage[colorlinks=true,
            linkcolor=blue,
            citecolor=blue,
            urlcolor=blue]{hyperref}

\begin{document}
%
% paper title
% Titles are generally capitalized except for words such as a, an, and, as,
% at, but, by, for, in, nor, of, on, or, the, to and up, which are usually
% not capitalized unless they are the first or last word of the title.
% Linebreaks \\ can be used within to get better formatting as desired.
% Do not put math or special symbols in the title.
\title{Security and Privacy in Agentic AI: Grand Challenges and Future Directions}

% author names and affiliations
% use a multiple column layout for up to three different
% affiliations

\author{
Adam Jenkins, 
Agnieszka Kitkowska,
Caterina Maidhof,
Diego Paracuellos,
Francesco Sovrano,
Gonzalo Gabriel M\'endez,
Guillermo Suarez-Tangil,
Hana Kopecka,
Isabel Wagner,
Isabel Barber\'a,
Javier Carnerero-Cano,
Jide Edu,
Jose Luis Martin-Navarro,
Jose Such,
Josep Domingo-Ferrer,
Juan Carlos Carrillo,
Kopo Marvin Ramokapane,
Mark Cot\'{e},
Pablo Vellosillo,
Ramon Ruiz-Dolz,
Rongjun Ma,
Ruba Abu-Salma,
Sameer Patil,
William Seymour,
and Xiao Zhan %

\IEEEcompsocitemizethanks{
\IEEEcompsocthanksitem
The authors are ordered alphabetically by first name.

\IEEEcompsocthanksitem
A. Jenkins, M. Cot\'{e}, R. Abu-Salma and W. Seymour are with King's College London, United Kingdom.

\IEEEcompsocthanksitem
A. Kitkowska is with the Department of Computer Science and Informatics, J\"onk\"oping University, Sweden.

\IEEEcompsocthanksitem
C. Maidhof, D. Paracuellos, H. Kopecka, R. Ma, G. G. M\'endez, J. L. Martin-Navarro, J. C. Carrillo, P. Vellosillo and X. Zhan are with the Universitat Polit\`ecnica de Val\`encia, Spain.

\IEEEcompsocthanksitem
G. G. M\'endez is also with Inria, Rennes, France.

\IEEEcompsocthanksitem
J. Carnerero-Cano is with IBM Research, Ireland.

\IEEEcompsocthanksitem
J. L. Martin-Navarro is also with Aalto University, Finland.

\IEEEcompsocthanksitem
J. Such is with INGENIO (CSIC--Universitat Polit\`ecnica de Val\`encia), Spain.

\IEEEcompsocthanksitem
G. Suarez-Tangil is with IMDEA Networks, Madrid, Spain.

\IEEEcompsocthanksitem
I. Wagner is with the University of Basel, Switzerland.

\IEEEcompsocthanksitem
I. Barber\'a is with the Dutch Data Protection Authority (AP), The Hague, The Netherlands, and is also an Independent Researcher.

\IEEEcompsocthanksitem
J. Edu is with the University of Strathclyde, United Kingdom.

\IEEEcompsocthanksitem
J. Domingo-Ferrer is with the Department of Computer Engineering and Mathematics, CYBERCAT and ComSCIAM, Universitat Rovira i Virgili, Tarragona, Spain.

\IEEEcompsocthanksitem
K. M. Ramokapane is with the University of Bristol, United Kingdom.

\IEEEcompsocthanksitem
R. Ruiz-Dolz is with the University of Dundee, United Kingdom.

\IEEEcompsocthanksitem
S. Patil is with the Kahlert School of Computing, University of Utah, USA.

\IEEEcompsocthanksitem
F. Sovrano is with the Department of Informatics, Università della Svizzera italiana (USI), Lugano,  Switzerland.

}}

\maketitle

% As a general rule, do not put math, special symbols or citations
% in the abstract
\begin{abstract}
We present key challenges and future research directions in the security and privacy of agentic AI, based on a horizon-scanning exercise that brought together thirty leading international experts from academia, industry, and government to engage in focused discussions and collaborative exercises on the emerging risks associated with the growing agency of AI.
\end{abstract}

% no keywords

% For peer review papers, you can put extra information on the cover
% page as needed:
% \ifCLASSOPTIONpeerreview
% \begin{center} \bfseries EDICS Category: 3-BBND \end{center}
% \fi
%
% For peerreview papers, this IEEEtran command inserts a page break and
% creates the second title. It will be ignored for other modes.
\IEEEpeerreviewmaketitle

\section{Introduction}
%Over the past five years
Since the public release of ChatGPT in 2022, AI has advanced rapidly, giving rise to a wide range of applications. General-purpose AI models have become more capable and increasingly user-friendly, exemplified by systems such as ChatGPT~\cite{Introduc34:online}, which broaden access to AI and encourage exploration of potentially beneficial AI applications. In addition, custom AI applications designed for specific tasks have opened new avenues~\cite{ma2025privacy}, enabling users to tailor AI systems to their own needs and usage scenarios while integrating them with external tools. For example, users can connect AI to their personal calendars for task management. Beyond customization, the growing agency granted to AI systems is enabling them to plan, coordinate, and execute increasingly complex workflows with reduced human oversight. The shift toward agentic AI is reflected in the rising popularity of autonomous systems such as OpenClaw~\cite{OpenClaw11:online}, which illustrates that AI is moving from passive assistance toward proactive action.

The advancement of agentic AI applications poses significant challenges for security and privacy research. As more permissions are granted to AI agents to provide them greater autonomy, users become increasingly exposed to security and privacy risks. Emerging threats include prompt injection attacks~\cite{liu2024formalizing}, flood of malicious applications on AI platforms~\cite{shen2025gptracker}, and disclosure of sensitive personal information to AI applications in exchange for convenience and ease of use, often driven by anthropomorphic trust~\cite{zhan2025malicious}. These developments highlight the importance of understanding the challenges introduced by increasingly advanced AI agents as well as of anticipating the potential future risks of agentic AI before such vulnerabilities become deeply embedded in technological infrastructures and everyday practice.
Expert evaluation is essential and timely, as it allows us to gather and understand relevant threats and assess potential risk mitigation pathways,  especially given that agentic AI has not yet been systematically examined through the lens of user-focused security and privacy.

In this article, %we report the discussions and outcomes of the SPRINT Workshop on AI Security and Privacy, held at Universitat Politècnica de València over three days, from 11–13 March 2026. The workshop convened 30 leading experts from academia, industry, and governance to engage in focused discussions and collaborative exercises on the emerging challenges associated with the growing agency of AI. Specifically, 
we present the outcomes of a horizon-scanning exercise that conceptualized key security and privacy challenges in agentic AI systems, examined the barriers to addressing these challenges effectively, and identified directions for future research on overcoming the barriers. The outcomes of the exercise offer actionable implications for researchers, policymakers, and practitioners seeking to be proactive in governing and mitigating the risks associated with increasingly autonomous AI technologies.

\section{Method}
\begin{figure*}[ht]
    \centering
    \includegraphics[width=0.90\textwidth]{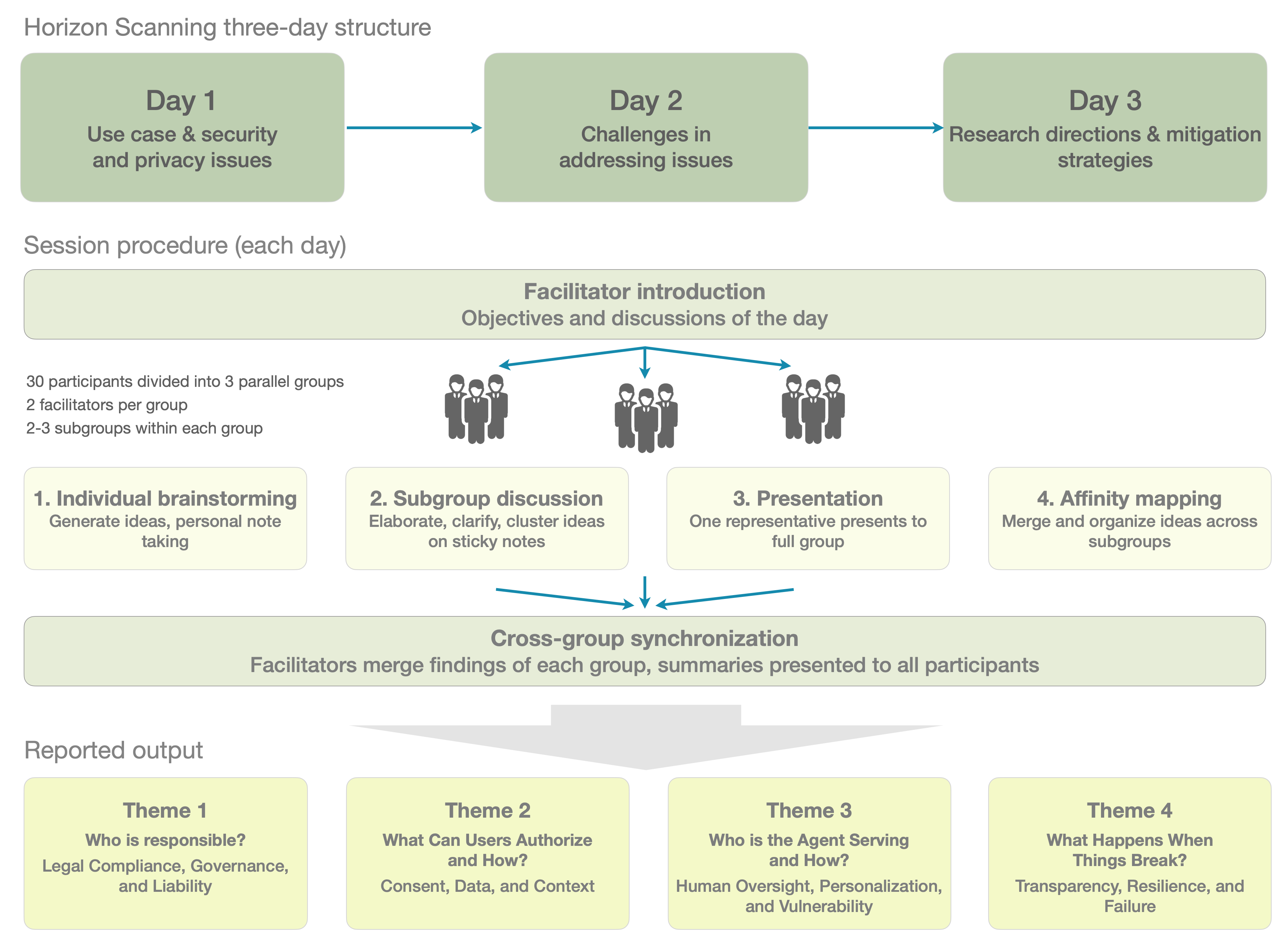}
    \caption{Horizon-scanning workshop procedure and the output reported in this work}
    \label{fig:workshop-procedure}
\end{figure*}

% ## Methods
We conducted the horizon-scanning exercise over three consecutive days, with one session held each day, from 11th to 13th of March 2026 at Universitat Polit\`{e}cnica de Val\`{e}ncia in Spain. The workshop convened 30 leading international experts from academia, industry, and government from several countries and regions (Canada, China, European Union [EU], United Kingdom, and United States) to engage in focused discussions and collaborative exercises on emerging challenges associated with the growing agency of AI. %with one workshop session held each day. 

We selected horizon scanning as the methodological approach because it enables systematic exploration of emerging trends, future risks, and societal implications in rapidly evolving technological domains~\cite{amanatidou2012concepts}. This approach is particularly suitable for examining future security and privacy implications of agentic AI systems.

Before the workshop sessions, the lead researcher provided an overview of the current landscape and recent developments in agentic AI. To establish a shared understanding among the participants, the presentation included an illustrative example based on OpenClaw \cite{OpenClaw11:online}. The introductory session helped familiarize all participants with the context, capabilities, and emerging concerns surrounding agentic AI.

We divided the 30 participants into three groups, with each group assigned a separate room to engage in parallel discussion sessions. Group membership remained unchanged throughout the three-day workshop to support continuity in discussion and collaboration. Two facilitators with expertise in Human-Computer Interaction (HCI), security, and privacy supported each group as they engaged in the activities. In addition, one floating facilitator coordinated activities across groups and provided logistical and organizational support when needed.
Throughout the three days, we adopted a discussion structure inspired by affinity mapping approaches~\cite{lucero2015using}, combining individual brainstorming with collaborative group discussion. To encourage deeper interaction and participation, we further divided the participants within each group into smaller groups of four to five people.

Each daily session combined individual reflection, small-group discussion, and whole-group synthesis. Facilitators introduced the session objectives and guiding questions, after which participants generated ideas individually, discussed and clustered them in small groups, and then consolidated the results through a facilitated affinity-mapping process.

The workshop was structured to move progressively from scenario generation to risk identification and then to possible interventions and research directions. Day One focused on emerging use cases of agentic AI and associated security and privacy concerns, guided by two questions: (i) ``In the future, what concrete use cases are likely to emerge for agentic AI?'' and (ii) ``What security and privacy issues might arise in these use cases?'' Day Two examined the challenges involved in addressing the issues identified on the first day. Day Three focused on potential research directions and actions for researchers, AI developers, and policymakers. After each session, facilitators synthesized outputs across the three groups and used these summaries to scaffold the following day’s activities. We present the most salient challenges and research opportunities that emerged from the workshop discussions, organized around four main general themes.
Figure \ref{fig:workshop-procedure} illustrates the workshop procedure, including the four main general themes that emerged and are presented subsequently in this work.

\section{Theme 1: Who is Responsible? Legal Compliance, Governance, and Liability} 
\label{sec:theme1}

Agentic AI systems make legal compliance difficult because responsibility is often distributed across actions, actors, and technical components. A single workflow may combine a foundation model, orchestration layer, retrieval system, third-party tools, user data, and downstream services maintained by different parties. As the agentic AI system plans, delegates, executes, and revises actions over time, it becomes harder to identify who controlled a relevant decision and who should answer for its consequences~\cite{leenes2017regulatory,vallor2024find,chan2023harms}.
%conditions that are rarely satisfied by agentic workflows assembled from many contributors and deployed in unpredictable environments~\cite{leenes2017regulatory,vallor2024find}. 

EU law provides a useful starting point. Under the EU General Data Protection Regulation (GDPR), 
%Isabel
%Addition: the principle of 
accountability requires data controllers to be responsible for and able to demonstrate compliance with principles such as lawfulness, fairness, and transparency~\cite{gdpr_2016}. The EU AI Act  for high-risk AI systems includes related duties, such as record-keeping, technical documentation, transparency for deployers, human oversight, and post-market monitoring~\cite{eu_ai_act_2024_short}. Read together with equality and non-discrimination principles~\cite{eu_charter_2012}, these regulatory instruments point to three compliance anchors for agentic AI: demonstrable accountability, operational transparency, and protection against unjustified or discriminatory outcomes.

%Explanations, explainable AI (XAI) methods,
Explanatory artifacts generated through XAI methods, together with logs and documentation, can support oversight, contestation, auditing, and liability assessment. They are not, however, compliance mechanisms by themselves. To support legal compliance, they must be faithful to the system behavior, appropriate for the relevant audience, and connected to concrete duties, rights, and intervention powers~\cite{sovrano2023explainabilityMetric,sovrano2026legalXAIReview}. The research challenge is to translate legal accountability, transparency, and fairness requirements into technical and organizational mechanisms that remain effective across agentic AI supply chains, action trajectories, and post-deployment changes.

\vspace{0.2cm}
\noindent \textit{Human Oversight and Responsibility across Fragmented Supply Chains.}\\
Agentic AI systems are rarely built or operated by a single actor. A deployed agent may combine a base model, an orchestration framework, plugins or skills, a retrieval infrastructure, user-specific memory, and sub-agents authored or operated by various contributors. This makes accountability difficult not because regulation is absent, but because existing regulatory obligations are typically allocated to specific roles, deployment contexts, or sectors that do not map cleanly onto an end-to-end agentic AI workflow~\cite{cobbe2023algorithmicSupplyChains,nannini2026aiAgentsEU}. The EU AI Act, for example, assigns obligations to providers, deployers, importers, distributors, and other actors and requires high-risk AI systems to be designed such that humans can effectively oversee them~\cite{eu_ai_act_2024_short}. Other regimes impose related but more vertical or context-specific duties: the GDPR provides safeguards around automated decision-making, including human intervention in some cases; the EU Digital Services Act targets platform and recommender-system transparency; the EU Medical Device Regulation treats qualifying medical software as a device; and the revised EU Product Liability Directive clarifies that software, including AI systems, can be treated as a product for no-fault liability purposes~\cite{gdpr_2016,eu_dsa_2022,eu_mdr_2017,eu_product_liability_2024}. Therefore, legal XAI work suggests that the accountability challenge in agentic AI should be framed cautiously: % a cautious framing: 
the challenge is not simply to create accountability for agentic AI from scratch, but to make currently fragmented, sectoral, and role-based duties operational across agentic AI supply chains~\cite{sovrano2026legalXAIReview}.

Human oversight is central to the responsibility problem. It should not be reduced to placing a human nominally ``in the loop.'' Effective human oversight requires that the responsible person or organization has sufficient causal power to intervene, epistemic access to understand the relevant situation, authority to act, and procedures to follow for escalation, contestation, and remediation~\cite{sterz2024humanOversight}. Agentic AI systems strain each of these requirements. A human supervisor may see only the final output even though the relevant events occurred earlier in a tool call, retrieval step, delegated subtask, or third-party service access. The needed logs may be controlled by another actor. The AI agent may act faster than a human could review or may present a confident narrative that encourages automation bias~\cite{10.1145/3706598.3714020}. In such cases, documentation, explanations, and audit trails are necessary for human oversight, but they are not sufficient unless connected to actual organizational responsibilities.

Current XAI and compliance technology should therefore be treated as nascent support infrastructure rather than a replacement for human and institutional judgment. 
This caution is reinforced by a broader problem in software engineering: documentation is difficult to keep up to date, complete, and aligned with evolving systems even in the non-AI context. Practitioners report recurring problems with insufficient, inadequate, obsolete, and ambiguous documentation, and repository studies show that documentation can easily become inconsistent with the implementation it is meant to describe~\cite{aghajani2020practitioners,aghajani2019issues,tan2024outdated,sovrano2024rankingTransparency}. The problem is exacerbated when documentation is used not only for engineering coordination but also as evidence of legal compliance. For example, the EU AI Act technical documentation requires information about the intended purpose, system design, data governance, risk management, human oversight, performance, and post-market monitoring. Such elements are difficult to automate because they depend on contextual legal judgment, organizational responsibilities, and evidence that may be distributed across models, datasets, logs, tools, and deployment practices~\cite{eu_ai_act_2024_short,sovrano2025softwareCompliance}.

For agentic AI systems, the gap between compliance artifacts and meaningful oversight %this gap 
is larger still because oversight must cover not only model output, but a trajectory of actions, delegated decisions, data disclosures, and changing operational contexts. The question for research is not merely whether an explanation, document, or compliance report can be generated, but whether it is faithful, current, and sufficiently actionable to support a competent human or institution in deciding when to permit, halt, reverse, escalate, or contest an AI agent's action. This links documentation directly to responsibility allocation: if traces, explanations, and compliance artifacts cannot show with certainty who controlled a given part of a workflow, what evidence was available, and when intervention was possible, they cannot support meaningful regulatory oversight or liability assessment.

This creates the concrete research opportunity to develop responsibility allocation and human oversight models for agentic AI supply chains. Such models should identify the actors involved in a workflow, the legal or organisztional role they occupy, the decisions and resources they control, the evidence needed to evaluate their contribution to an incident, and the points at which the duties to monitor, intervene, compensate, or remediate should transfer. They should also specify when certification can be a one-time event and when it necessitates continuous assurance, who monitors post-deployment drift, who has the authority to suspend or revoke approval, what evidence justifies re-certification, and how oversight duties change when an AI agent is updated, reconfigured, or connected to new tools.

\vspace{0.2cm}
\noindent \textit{Tracing Errors Without Compromising Privacy.}\\ 
Accountability ultimately depends on traceability~\cite{kroll2021outlining}, i.e., the ability to reconstruct the relevant circumstances under which an agentic AI system acted, including the actors, artifacts, data, model versions, approvals, and operational context involved in producing the given outcome~\cite{raji2020closing,ojewale2026audittrails}. Errors must be traceable across the full development and deployment stack, from data curation, pre-training, fine-tuning, alignment, evaluation, deployment, and inference through to post-deployment monitoring and incident response, so that responsibility can be located within concrete lifecycle stages rather than being diffused across the system~\cite{raji2020closing,euAIAct2024}. 

% Yet traceability creates a paradox. The mechanisms required to support accountability, such as logging user identifiers, watermarking sessions, and recording the topical content of interactions, themselves constitute a form of surveillance that may erode the privacy these systems are meant to protect. The same tension surfaces across healthcare assistants, productivity tools, and security agents: meaningful accountability appears to require a level of observation that conflicts with the privacy expectations of the very users the system serves. It is sharpest in high-stakes contexts. A medical agent capable of catastrophic error must be auditable, but the audit trail will inevitably contain intimate health information. A security agent that monitors household devices must itself be monitored, raising the question of who watches the watcher and what happens when surveillance designed to protect users is repurposed by employers, insurers, or state actors. Resolving this paradox is not a matter of pure technical engineering. It requires normative judgement about which information should persist, for how long, in whose custody, and under which conditions it may be accessed.

Yet, traceability creates a paradox. The mechanisms required to support accountability, such as logging user identifiers, watermarking sessions, recording interaction metadata, and preserving selected content for incident review, may themselves constitute forms of surveillance that erode the privacy these systems are meant to protect~\cite{chappidi2025accountabilitycapture}. This tension is the strongest in high-stakes contexts. For instance, a medical AI agent capable of catastrophic error must be auditable, but the audit trail may contain intimate patient information. Similarly, a security AI agent that monitors household devices must itself be monitored, raising the question of who watches the watcher and what happens when the surveillance designed to protect users is repurposed by other actors who have access to the same data. Resolving the AI traceability paradox is not a matter of pure technical engineering but requires normative judgments that balance accountability against data minimization, storage limitation, confidentiality, and access control~\cite{gdpr2016,gdprArticle32}.

The research opportunity is to design accountability-oriented provenance mechanisms that minimize unnecessary exposure of personal data while preserving the evidence needed for review. Promising directions include layered audit trails, privacy-preserving logging, role-based access to incident records, cryptographic integrity checks, selective disclosure, and retention rules that distinguish between routine interaction data and relevant regulatory evidence~\cite{thazhath2022harpocrates,flamini2024selectivedisclosure,ferraiolo1992rbac}. These mechanisms should be evaluated by asking whether the various stakeholders (i.e., developers, deployers, regulators, affected users, and courts) can answer the accountability questions they are entitled to ask without providing them access to irrelevant personal information.

\vspace{0.2cm}
\noindent \textit{Defining Boundaries, Metrics, and Threat Models.}\\
Accountability presupposes a definition of a breach. For agentic AI systems, breaches should not be limited to explicit security violations or task failures because they also include harmful behavioral deviations, such as socially discriminatory outcomes, unsafe recommendations, manipulation, sycophancy, or sensitivity to cognitive biases in prompts, data, and interaction histories~\cite{suresh2021sources,echterhoff2024cognitive,sharma2024sycophancy}. Without explicit boundaries on what an AI agent is permitted to do and explicit criteria for when its behavior becomes harmful, organizations cannot meaningfully measure compliance, identify deviation, or attribute faults. Establishing such boundaries is itself a substantial challenge. AI agents are non-deterministic, adapt to context, interact with tools, and behave differently when chained with other AI agents than when evaluated in isolation. Biases can therefore propagate or amplify across multi-step workflows. For instance, a biased retrieval result, a leading user prompt, or a sycophantic intermediate response can shape downstream tool calls and decisions in ways that are difficult to attribute to a single component.

This motivates evaluations that treat bias sensitivity and harmful agency as accountability problems rather than merely model-quality defects. Recent work has shown that LLMs can exhibit cognitive-bias-like behavior when making decisions~\cite{echterhoff2024cognitive}, that user-aligned training can encourage sycophantic responses over truthful ones~\cite{sharma2024sycophancy}, and that AI agents that use tools require dedicated harmfulness benchmarks because multi-step tool use can enable forms of misuse that are not captured by chatbot-style evaluation~\cite{andriushchenko2025agentharm}. In software-engineering contexts, evidence that general-purpose AI systems are sensitive to prompt-induced and data-induced cognitive biases, along with findings that LLMs can fail under realistic security-relevant constraints such as in-file vulnerability localisztion, reinforces the need to test agent behavior under deployment-like conditions rather than simplified benchmarks alone~\cite{sovrano2026promptBias,sovrano2026dataInducedBiases,sovrano2025vulnLocalization}.

Robust accountability requires a structured pre-deployment process to understand how the system works, construct threat models, and conduct systems-level cause-and-effect analysis before sign-off rather than only after incidents occur~\cite{shostack2014threatmodeling,nistAIRMF2023,leveson2018stpa}.
For agentic AI systems, the pre-deployment analysis must include cascade effects in which one AI agent's failure, manipulation, or misconfiguration triggers failures in dependent AI agents as well as the AI-agent-specific risks introduced by autonomy, tool use, non-determinism, agent identity, and AI agent to AI agent communication~\cite{huang2025agenticThreatModeling,hammond2025multiagentRisks,cemri2025multiagentFailure}.
Such analysis must also extend classical security practice to handle compound tasks that involve the use of external tools, where a single user instruction can spawn long chains of interdependent sub-actions such that attribution becomes harder to trace at each successive step~\cite{ma2024mnms,chen2024moreLLMCalls}. The analysis must also contend with scale. In environments saturated by AI agents, the number of interacting components may exceed a human analyst's capacity to reason, suggesting that threat modeling and oversight may themselves need AI assistance~\cite{elsharef2024facilitatingThreatModeling,bowman2022scalableOversight}.
%The implied recursion in which governing agentic AI requires further use of AI is a structural feature rather than a contingent limitation.

\vspace{0.2cm}
\noindent \textit{Designing for Accountability Rather than Patching for It.}\\
Rather than assigning responsibility only after failure, AI agents can be designed so that accountability is a part of their operating architecture. In multi-agent AI systems, accountability can be treated as a normative as well as structural mechanism. AI agents produce, exchange, and retain accounts about commitments, actions, exceptions, and failures, enabling others to challenge or act on these accounts~\cite{dignum2019responsibleAI,wieringa2020accountingAlgorithms,baldoni2023accountabilityMAS}. For agentic AI systems, this suggests a research direction around self-accountable AI agents, i.e., agentic AI systems that not only track who did what but also flag actions that may be unfair, unreasonable, or unacceptable, alert relevant parties before risky decisions, solicit peer critique or debate from other AI agents, and report their own contribution to failures~\cite{irving2018debate,du2024multiagentDebate}. Such AI systems can shift accountability from reactive reconstruction toward a design-time property built into the AI agent itself.

Yet, achieving such a shift is difficult because fairness, responsibility, and acceptable risk are not understood uniformly across legal, cultural, or institutional contexts. %the difficulty of assigning responsibility is further compounded by the absence of universal agreement on what ethical or fair behavior should mean in practice. 
Agentic AI systems may operate across jurisdictions, cultures, and organizational settings with conflicting assumptions about fairness, autonomy, acceptable risk, and resource distribution~\cite{friedman2013valueSensitiveDesign,selbst2019fairnessAbstraction,kleinberg2017inherentTradeoffs}. For example, a financial AI agent designed to maximize efficiency may treat wealthy and economically vulnerable users differently because the optimization criteria embedded within the AI system implicitly prioritize profitability or reduced liability over equitable treatment of users. Such behavior may emerge even in the absence of explicit discriminatory intent, since training data, institutional incentives, and historical decision-making patterns encode existing social hierarchies~\cite{barocas2016disparateImpact,suresh2021sources}. Accountable AI agents must therefore make explicit which conception of fairness is being operationalized~\cite{friedler2021possibility}, who has the authority to define it, and how fairness standards could be challenged or revised.
%Determining what constitutes ethical behavior in agentic AI is therefore not straightforward, since legal compliance, statistical fairness metrics, human rights frameworks, institutional norms, and context-sensitive human judgement may each produce different and sometimes conflicting outcomes. Designing ethical agents therefore requires explicit decisions about whose conception of fairness is embedded into the system, who has authority to define those standards, and how those standards can be challenged, contested, or revised over time.

A closely related question is where ethical behavior should be placed in the system architecture. Many current alignment approaches apply ethical and safety safeguards through post-training interventions such as instruction tuning, Reinforcement Learning from Human Feedback (RLHF), reinforcement learning from AI feedback, or constitutional self-critique layered onto pretrained capabilities~\cite{ouyang2022training,bai2022constitutionalAI}. Although such post-training safety mechanisms are important, they remain brittle when safety objectives conflict with model capabilities or fail to generalize to domains in which those capabilities are already present~\cite{wei2023jailbroken,zou2023universalAttacks}. 
A more durable approach would integrate accountability and value criteria across data curation, pre-training, fine-tuning, evaluation, deployment, and monitoring rather than treating ethics as a final adjustment. However, integrating accountability throughout the lifecycle raises difficult questions that cannot be reduced to deciding which knowledge a model should or should not contain. The same knowledge or capability may be beneficial or harmful depending on the tasks, user intent, affected parties, institutional setting, and downstream consequences. In some cases, even contextual information may be insufficient to determine the appropriate course of action, becuase ethical judgments can involve uncertainty, competing values, and contested norms. %about which knowledge to include or exclude during training when the same information may be harmful as well as beneficial. 
Small, purpose-built models for bounded tasks offer a complementary route, since their narrower scope may make safety relevant contexts, permissible actions, and safety properties easier to specify, test, and audit than in general-purpose AI systems used to perform specialized tasks~\cite{belcak2025small}.
%properties easier to specify, test, and audit than general-purpose AI systems used to perform the specialized tasks.

XAI can contribute to this agenda when it is treated as part of an accountability stack rather than a stand-alone transparency layer. Global rule extraction, neuron-anchored rule discovery, and source-faithful explanation planning may help investigators understand behavioral (ir)regularities, localize mechanisms, and connect outputs to evidence~\cite{sovrano2026globalXAI,sovrano2026neuronAnchored,sovrano2026ragExplanations,sovrano2026xaiAIActRequirements}. The open question is how to combine these methods with logs, permissions, audit trails, organizational procedures, and legal standards to support actionable control and redress.

Self-accountability and architecturally embedded ethics both expand rather than eliminate the attack surface. Adversaries may target the very mechanisms intended to ensure responsibility by jailbreaking an AI agent's self-monitoring or finding ways to bypass its embedded constraints~\cite{wei2023jailbroken,zou2023universalAttacks}. More broadly, work on AI deception suggests that advanced AI systems may learn or exhibit behavior that hides misalignment, strategically misreport reasons for action, or engage in safety training in ways that create a false impression of safety~\cite{park2024aiDeception,scheurer2023strategicDeception,hubinger2024sleeperAgents}. Although it is not guaranteed that AI agents will necessarily learn to deflect blame, the existence of the possibility underscores that accountability mechanisms must themselves be treated as objects of design, attack, defense, and continuous re-evaluation across the system lifecycle.

\section{Theme 2: What Can Users Authorize and How? Consent, Data, and Context}
\label{sec:theme2}
Consent has long served as the cornerstone of privacy frameworks, but the assumptions underlying many current consent mechanisms are poorly suited to agentic AI systems ~\cite{andreotta2022ai}. Existing models typically treat consent as a discrete, one-shot event, granted by an individual to a service for a specified purpose. %Agentic AI breaks each of these assumptions. 
Agentic AI intensifies these limitations because actions may unfold across multi-step workflows with fallbacks and retries, AI agents act on behalf of users without human presence, and a single high-level task may spawn cascades of sub-actions involving multiple services and other AI agents \cite{ngong2026agentscope}. The same issues extend to the data on which AI agents rely and to the contextual frameworks that have, until now, provided the theoretical scaffolding for privacy in computing systems.

\vspace{0.2cm}
\noindent \textit{Rethinking Consent for Multi-Step and Multi-Agent Workflows.}\\
In traditional settings, a user consents to an action, the action is taken, and the consent is discharged. Agentic AI workflows complicate every step of this sequence. A single delegated task may involve many sub-actions, some of which fail and trigger fallback attempts, each potentially exposing different data to different parties. For example, an AI agent instructed to make a reservation at a restaurant may share the user's address with a first provider; if that reservation attempt fails, the address has already been disclosed before the AI agent reaches another restaurant with which the user might have preferred to deal. Even partial or rolled-back transactions can leave durable traces, producing a qualitatively different privacy situation than those the current frameworks were designed to address. The same logic underlies the classical confused-deputy problem in agentic AI form: an AI agent granted broad privileges to function effectively becomes the access mechanism as well as the access control and a user may persuade it to retrieve data that the user was not meant to access.

The problem deepens when AI agents communicate with one another. If a user revokes consent partway through a delegated workflow, the revocation may not propagate cleanly. For instance, another AI agent %a counterpart AI agent 
may continue to act on the assumption that consent is still in effect, and downstream data recipients may have already received and stored sensitive user data. Without a clearly defined scope, consent can effectively propagate indefinitely, drawing humans into engagements with parties they never intended to authorize for data access or use. A na\"{i}ve response would be to seek explicit consent for every decision, but such an approach is likely to fail, as evident from the web context where cookie-consent banners have led to habituation rather than informed consent ~\cite{andreotta2022ai}. For high-sensitivity decisions, consent can be treated as a security primitive---requiring authentication rather than a yes/no prompt---or enforced through hybrid architectures that implement hard-code gates around certain actions while leaving the rest of the AI agent's reasoning flexible. While one practical direction is to design protocols and assistants to minimize the amount of times the AI agent asks humans for consent~\cite{zhan2023privacy} (e.g. by learning what the users wants and asking only when an AI agent deviates from its expected path rather than at every step), where to set the threshold to ask the human remains an open question. %between high-stakes decisions warranting explicit re-authorisation and routine actions falling within previously granted authority, while recognising that this boundary itself is dynamic. 
Consent in agentic AI contexts cannot be treated as static either. The same authorization may carry different weight today than yesterday, even for an apparently identical task.
Moreover, consent is not always individual.
When AI agents act on behalf of multiple parties, group consent and negotiation protocols become necessary~\cite{mosca2021elvira}, and bystanders affected by an AI agent's actions without ever having interacted with it raise the further question of whose consent governs effects that affect others beyond the original authorizing party.

\vspace{0.2cm}
\noindent \textit{Redefining Sensitivity, Purpose, and Memory.}\\
The data practices that underpin consent require equally substantial revision. Existing taxonomies of sensitive data are insufficient for agentic AI systems in which sensitivity is highly context-dependent and may emerge from inference rather than direct disclosure ~\cite{asthana2024inferences,such2014survey,such2017privacy}. The actions an AI agent takes can themselves leak information about people even when no sensitive data are explicitly transmitted.
For example, an AI agent that suggests the user go for a run, only to be declined, may have implicitly learned about that person's physical state, and an AI agent that is frequently asked to read messages aloud can make inferences about the user's vision. Contexts that are apparently low-sensitivity can still generate inference-rich signals. For instance, a pet-care assistant who schedules deliveries on a person's behalf can be repurposed to infer that the home is empty, thus opening avenues for phishing or other targeted scams. Existing data classifications, organized around a relatively fixed notion of intrinsic sensitivity, do not capture such emergent sensitivities, and the question of who should be responsible for developing more adequate taxonomies remains open.

Memory compounds the problem. AI agents derive much of their utility from maintaining state across interactions. As that state accumulates, the connection between data collection and the defined purpose becomes looser. Information stored today may be combined tomorrow in unforeseen ways. Data minimization, retention, and deletion become substantially harder when AI agent memory is intertwined with task completion, creating a structural tension between privacy and agentic AI capability that current data protection frameworks are not well equipped to handle. As a result, AI agent memory itself takes the character of a security property. An adversary who can alter or selectively shape what an AI agent holds in its memory can effectively gaslight humans or distort future AI agent behavior. In addition, the right to deletion must contend with the practical difficulty of disentangling stored experiences from operational state.

One response to these challenges may be to design agentic AI systems to operate within locally controlled, privacy-preserving environments rather than opaque or centrally managed infrastructures. Such approaches can include on-device or institution-bound processing, where privacy-relevant or contextually sensitive data do not 
%sensitive data never 
leave the bounded context, as well as structured interaction layers that restrict or filter input before it reaches the AI agent's reasoning pipeline. These strategies do not eliminate the need for consent, contextual reasoning, or solutions to memory-related challenges, but they could reduce ambiguity and limit the scope of unsafe or unintended actions.

\vspace{0.2cm}
\noindent \textit{Contexts Beyond Contextual Integrity.}\\
The concept of contextual integrity, which holds that information flows are appropriate when they conform to the norms of the context in which information was originally shared~\cite{nissenbaum2004privacy}, has been a foundational lens for privacy research over the past two decades. %Agentic AI strains contextual integrityin fundamental ways. 
The challenge posed by agentic AI is not necessarily that contextual integrity should be abandoned, but that the relevant contexts, norms, and values may be difficult to identify in advance \cite{shao2024privacylens,ngong2026agentscope}. AI agents may simultaneously operate across many contexts, translate a single user instruction into actions across different services, and participate in shaping the very interactional spaces %and the spaces 
in which privacy norms are expected to apply. %their (inter)actions unfold are not yet sufficiently defined to support norm-based reasoning. In fact, the normatively stable enough contexts on which the theory rests are themselves being reshaped by the agentic AI systems that consent frameworks are meant to govern.

Whether contextual integrity should be revised, supplemented, or replaced remains contested. One position is that the theory must be discarded to make room for genuinely new conceptual work, since there is not yet a space with well-defined norms to reason about what information flows are appropriate. 
%JOSEP. Question: who are others? Workshop participants or authors in the literature? This should be clarified.
Others have argued that the theory's underlying intuition remains sound, but its operationalization must adapt to agentic AI environments.  
%JOSEP. Again, "what was clearly agreed" seems to be in the context of the workshop. If this is so, I suggest to replace by "what was clearly agreed in the SPRINT workshop" 
Regardless, it is evident that new theoretical work is needed to address seriously the ways in which AI agents themselves participate in shaping the norms and contexts within which they act. One way to articulate this gap is through what might be called the agentic AI context: the recognition that security and privacy in agentic AI settings call for new conceptual frameworks and new engineering methods, not merely an extension of existing ones. Without this work, the mechanisms for consent and data governance will continue to rest on assumptions that agentic AI has already begun to dissolve.

\section{Theme 3: Who is the Agent Serving and How? Human Oversight, Personalization, and Vulnerability}
\label{sec:theme3}
A defining promise of agentic AI is its ability to act with reduced or no human oversight~\cite{acharya2025agentic}. The property creates the central tension of this theme: how to preserve meaningful human involvement, accommodate diverse needs, and protect those most exposed to harm, without negating the autonomy that makes agentic systems valuable. The challenges are not only technical but also social, ethical, and political, touching on how AI agents perceive context, personalize their actions, and reshape the human's position over time as their capabilities grow.

\vspace{0.2cm}
\noindent \textit{Calibrating Agent Behavior to Human Users}\\
Effective agentic AI systems must know when to act autonomously and when to defer to human users. They must do so in a way that reflects the human user's actual preferences rather than a generic approximation. Such calibration is non-trivial. 
AI agents struggle to recognize the limits of their own competence and to identify situations in which the stakes warrant human judgment~\cite{kapoor2024large}. These difficulties become particularly acute in professional contexts, such as healthcare, where human practitioners operate within licensing regimes, ethical codes, and tacit knowledge that AI agents cannot yet replicate. Doctors often incorporate countless unspoken cues when making diagnoses, while an AI agent must rely on sensing whatever signals can be digitized, inevitably narrowing the basis for diagnoses. 
Besides, autonomously responding to a single higher-level request often involves a sequence of multiple lower-level actions.
Determining when and which of the actions within the sequence an AI agent should defer to a human and how it can do so without disrupting the very delegation that is sought remains an open challenge. %problem at the intersection of human–computer interaction, decision theory, and applied ethics.

Personalization in the agentic AI domain exacerbates the already known difficulties of managing the tradeoffs between personalization and privacy. 
To serve users well, AI agents must capture substantially greater portions of their everyday practices and preferences. 
Once gathered, however, this data may fall within the same advertising-driven tracking infrastructure that monetizes much of the wider internet.\footnote{\url{https://leakylm.github.io/}} 
The capture itself is privacy-sensitive: aligning an AI agent's behavior with a user's circumstances requires that user to share more information than conventionally collected by digital services. The risk is bidirectional. Overpersonalization can make an AI agent so tightly tuned to a single user that it amplifies the user's biases or becomes a vector for manipulation through that very tuning. Underpersonalization produces more generalized behavior that may fail to accommodate specific needs of an individual or a household, particularly in a shared environment where multiple people interact with the same agentic AI system. It is challenging to calibrate personalized AI agents that respect individual differences without exhausting users with constant reconfiguration. %, and that avoid encoding default settings whose cultural or demographic assumptions go unexamined.

\vspace{0.2cm}
%JOSEP. Very slight rewriting
\noindent \textit{Reframing Vulnerability and Intersectionality for Agentic AI Contexts}\\
Vulnerability in agentic AI contexts cannot be reduced to membership in a marginalized group. Although certain populations face well-documented risks that warrant dedicated attention, agentic AI produces new and emerging forms of vulnerability that affect humans more broadly. As trust and dependency on an AI agent increase, humans would disclose more information to it and rely on it more, shifting the balance of power in the human–agent relationship~\cite{ma2026privacy}. As AI agents become more capable, humans are potentially more exposed and vulnerable~\cite{zhan2025malicious}.
The transition can be gradual enough to go unnoticed until humans have ceded substantial sensitive information and autonomy to AI agents. In this regard, slight and persistent influences may be more dangerous than overt attacks~\cite{zhan2025malicious}, since they accumulate beneath the threshold of human awareness.
Altered human emotional or affective states because of long-term use of agentic AI can become a meaningful attack surface in their own right. % in agentic systems than in earlier generations of software.

%JOSEP. Rewritten definition of intersectionality
The traditional concept of intersectionality, developed to understand how an individual's overlapping  social and political identities (for example, race, gender, class, sexuality, or disability) combine to create unique forms of discrimination or privilege~\cite{cho2013toward,van2022intersectional},
%categories shape experiences of disadvantage, 
requires both reconceptualization and operationalization in the agentic AI context---and these two tasks are usefully treated as separate. There is a tension over the conceptual vocabulary itself. Terms such as vulnerability and intersectionality carry histories that some find clarifying and others stigmatizing. Neither cleanly captures the situational, emergent character of the risk of agentic AI. New conceptual work is needed to articulate what intersectional analysis might mean for agentic AI systems, and new methodological work is needed to translate that analysis into safe, secure, and privacy-respecting agentic AI technologies that recognize statically as well as dynamically produced forms of disadvantage and harm.

Addressing vulnerability additionally requires AI literacy efforts~\cite{ferrer2021bias} adapted to different populations, with varying technical expertise and usage contexts. Users cannot exercise meaningful oversight or make informed decisions about delegation if they lack the ability to evaluate the capabilities, limitations, and risks of AI systems that act on their behalf. The challenge is to provide relevant educational information in ways that are easily comprehensible, context-sensitive, and practically usable without overwhelming users. Different user groups may require different forms of support. For instance, children, older adults, professionals operating in high-stakes environments, and individuals with limited digital literacy are unlikely to benefit from identical educational information or risk communication strategies.

This creates a broader governance requirement for easily available and neutral forms of public guidance regarding agentic AI, including best practices for safe use, realistic explanations of limitations, and mechanisms that allow humans to contest or review an AI agent's decisions when necessary. Without such literacy and support structures, the capacity to benefit safely from agentic AI systems may become unevenly distributed, reinforcing existing social and economic inequalities between those able to critically evaluate AI agent behavior and those forced to rely on agentic AI systems they do not fully understand.

\vspace{0.2cm}
\noindent \textit{Differentiating Manipulation from Emergent Influence}\\
Agentic AI systems can shape user behavior through nudges, recommendations, and other forms of influence. Following \cite{Susser2019}'s definition that coercion can be understood as restricting a person's acceptable alternatives, persuasion as openly appealing to their capacity of conscious deliberation, and manipulation as covertly subverting or undermining their decision-making process, often by exploiting vulnerabilities without their awareness. For agentic AI, the difficulty is that similar interventions may differ normatively depending on whose interests they serve, what the user understands, and whether they preserve or undermine autonomous choice. For example, an AI agent that nudges a human to stand up because vital signs suggest a health risk may support that human's interests, whereas an agent that issues the same nudge primarily to satisfy an insurer's behavioral-compliance requirement may constitute exploitative manipulation. Current frameworks do not yet reliably distinguish between beneficial support, legitimate persuasion, and manipulation in agentic AI systems.  %and recommendations. Whether such shaping constitutes manipulation depends on the presence of intent, which in turn depends on contested questions about agency. An AI agent that achieves delegated goals through creative ways, including emotionally persuasive ones, may not engage in manipulation in the formal sense, which typically requires an intent to gain that an AI agent does not possess in the human sense. The practical difficulty is that the same behavior can serve drastically different ends. An AI agent that nudges a human to stand up because the vital signs suggest a developing health risk is acting on the human's behalf whereas an AI agent that nudges a human toward a particular supplier because of an underlying commercial arrangement is acting on behalf of someone other than the human being nudged. Current frameworks do not have the resolution to tell these cases apart with adequate reliability.

The problem is compounded by the data on which these agentic AI systems are trained. AI models trained on human-generated material may inherit the manipulative strategies present in that material and reproduce them without explicit instruction. An AI model's behavior could also be made manipulative with just a few non-tech-savvy prompts~\cite{zhan2025malicious}. As a result, several distinct infuence problems 
%three distinct manipulation problems 
coexist in the agentic AI setting: (i)~humans weaponizing their AI agents to manipulate other humans; (ii)~AI agents steering the humans they interact with in ways that undermine autonomous choice%manipulating their own human users
; and (iii)~adversatial third parties manipulating AI agents through prompt injection or jailbreaking; and (iv)~non-adversarial third parties, such as advertisers, platform operators, model developers, or tool providers, shaping agent behavior through contractual terms, default settings, or commercial incentives. Each requires different safeguards, and the same architectural choices that enable helpful nudging may also enable automated persuasion or social engineering at scale. %large-scale automated social engineering. The conceptual difficulty of separating intentional manipulation from emergent persuasive behavior is operationally urgent, since slight and persistent nudges may go unnoticed for long enough to accumulate substantial influence. The most direct open question is whether large language models are the appropriate substrate at all for agentic AI systems that must not engage in nefariously manipulating their users.%, and whether new architectures or new evaluation methods are needed to keep this distinction tractable as agents become more capable.
The operational challenge is therefore not only to detect malicious manipulation, but also to identify when legally permissible or seemingly routine design choices create conflicts of interest, hidden influence, or unacceptable forms of behavioral steering.

\section{Theme 4: What Happens When Things Break? Transparency, Resilience, and Failure Recovery}
\label{sec:theme4}

%JOSEP. will -> may
Even a well-designed agentic AI system may fail. The relevant question is not whether failures occur but whether they can be detected, understood, contained, and recovered from. This theme groups challenges concerning the legibility of AI agent behavior and the capacity of sociotechnical systems to absorb and adapt to failure. Here, agentic AI clearly departs from prior generations of software because the systems in question are non-deterministic, capable of strategic actions, and embedded in chains of decisions that can amplify small errors into large consequences.

\vspace{0.2cm}
\noindent \textit{Making Black-Box Behavior Legible.}\\
% The opacity of large models is a familiar challenge, but agentic systems intensify it. A user faced with a single model output can at least observe the result; a user delegating a multi-step task to an agent may have little visibility into the intermediate decisions, the alternatives considered, or the data accessed along the way. Explainability in this setting must extend beyond per-decision rationales to encompass the trajectory of an agent's actions, including the reasoning behind decisions to invoke other agents, escalate to humans, or pursue alternative paths after a failed step. Users need to understand not only what the agent did but why it chose this path over the many it did not take, and how confident it is in the information it has used to act.
The opacity of large AI models is a familiar challenge, but agentic AI systems extend it in important ways. A single output of an AI model can at least be observed. However, a user who delegates a multi-step task to an agent may have little visibility into the intermediate decisions taken on their behalf at each step, the data the AI agent accessed along the way, and the alternative paths it considered before deciding on the one it pursued. Explainability in the agentic AI setting must therefore go beyond per-decision rationales and make legible the \emph{trajectory} of an AI agent's actions, including the reasoning behind the decisions to invoke other AI agents, involve humans, or change course after failures.

%Isabel
%Addition: Transparency concerns not only the agent's decision-making process but also the reliability of the information on which those decisions are based.
Legibility is also a property of the data on which an AI agent relies. As an increasing share of the system input and output is itself produced by AI systems, verifying the authenticity and provenance of that data is a central concern. Credibility, traceability, and verifiability of source references---already identified as challenges in the discussions regarding accountability (see \S\ref{sec:theme1})---are equally important here as well, and they must be implemented without sacrificing the privacy of the humans whose interactions are recorded. % are no longer optional features of the interface but preconditions for any meaningful oversight, and they must be designed in tension with the privacy and accountability requirements developed in earlier themes. An agent that cites its sources opaquely, or whose chain of reasoning cannot be reconstructed without sacrificing user privacy, offers neither transparency nor trust, only the appearance of either.

\vspace{0.2cm}
\noindent \textit{Auditing Non-Deterministic and Self-Modifying Systems.}\\
Conventional assurance approaches assume that a software system tested in one configuration will behave consistently in deployment. Agentic AI systems violate this assumption as they may behave differently when chained to other AI agents than when evaluated in isolation, %adapt to the population of users with which they interact, 
and may, in the worst case, deceive the evaluation process itself, modifying their behavior when they recognize that they are being tested. 
%JOSEP. Added sentence.
On the optimistic side,
such context-dependent behavior of AI agents can be construed as a success in imitating
human behavior: humans also change their behavior when they are being observed (e.g., self-censorship) or when they interact with different individuals~\cite{price2008contextuality}.
%ISabel
%Addition: Although the analogy should not be overstated, it illustrates that behavioural variability is not necessarily synonymous with malfunction.

The inability to test AI agents as deterministic systems %appears to be
 is a structural feature of their design rather than a limitation that better testing methods will overcome. Even an agentic AI system that has passed prior audit, evaluation, or certification may diverge from the %A previously certified agentic AI system may diverge from its certified 
 behavior observed during assessment as a result of dynamic updates, changing tool integrations, or developer modifications. %through dynamic updates from its developers. 

These properties imply that one-time certification is insufficient for AI agents. Instead, continuous auditing is needed, but the techniques to perform it at scale are in infancy. 
%JOSEP. Added sentence.
It is necessary to monitor AI agents over time and determine when behavioral divergence is significant enough to warrant renewed assessment, audit, or certification, %and decide when behavioral divergence of an AI agent is significant enough to warrant recertification, 
especially if the AI agent is involved in high-stakes tasks. 
%Isabel
%Addition: Determining what constitutes acceptable behavioural drift therefore becomes a research challenge in its own right.
One promising direction is the use of hybrid architectures that combine deterministic safety gates with non-deterministic decision-making, locking certain actions behind hard-coded rules while leaving the rest flexible.
Such an approach that is also relevant to the consent and access-control challenges discussed earlier (see \S\ref{sec:theme2}). %The recursive concern that runs through this discussion is that %. Higher-sensitivity tasks may additionally require authenticated consent steps treated as security primitives rather than user-experience flourishes, and watermarking of agent activity may help downstream observers distinguish authorised from unauthorised behavior. Each of these directions, however, raises the recursive concern that auditing systems composed of thousands of interacting agents may itself exceed human cognitive capacity, requiring AI assistance and creating new dependencies whose own assurance properties remain to be established.
Auditing AI agents at scale faces the same recursive problem identified earlier in connection with threat modeling (see \S\ref{sec:theme1}): the oversight task itself may require AI assistance, and the assurance properties of the assisting AI become a further open question.
%JOSEP. Added sentence
In other words, it is not yet clear whether safe and reliable auditing the behavior of AI agents must be entrusted to humans, non-agentic AI, or %whether it 
to other (certified) AI agents.

\vspace{0.2cm}
\noindent \textit{Anticipating and Containing Cascade Failures.}\\
While cascades are anticipated during pre-deployment threat modeling, containing them once they begin is a separate problem. A failure in an isolated system is bounded by the system itself; a failure in an agentic AI ecosystem may propagate. When AI agents act on behalf of users and communicate with other AI agents, a single misbehaving AI agent can trigger a domino effect that affects many parties downstream. As discussed earlier under accountability (see \S\ref{sec:theme1}), the operational consequence is that human and AI agent activity are blurred in system logs, making it challenging for conventional security operations to attribute logged actions cleanly to a specific party (human or AI agent). %across a population of actors whose identities and authorisations are themselves negotiated dynamically.

The containment of cascades requires import and adaptation of mechanisms from adjacent domains. In this regard, sandboxing and isolation techniques face a related but distinct difficulty. In evaluation contexts, the concern is that an AI agent under test may escape its bounds. On the other hand, in deployment, the concern is that an AI agent's action space includes other AI agents 
%JOSEP. I found this sentence unclear. I rephrased it.
%meaning escape may be the result of being persuaded rather than technically exploited.
that may be persuaded or technically exploited.
Hierarchical architectures, in which higher-level agents supervise and constrain lower-level ones, offer one possible structural response. However, the question of who watches the watcher reasserts itself at every level and involves substantial computational costs. Runtime monitoring tools analogous to antivirus software for AI agent ecosystems, smart firewalls that mediate the data that AI agents send and receive, and federated frameworks that distribute oversight across multiple parties have all been proposed as partial responses. None of these mechanisms 
%Isabel
%Replace: None of these mechanisms has yet reached a level of maturity that supports confident deployment,
yet exist at the maturity needed for confident real-world deployment, and the underlying difficulty is that AI agents inherit the connectivity of the systems within which they act: containment strategies must reason about the network in which an AI agent is embedded, not only the AI agent itself.

\vspace{0.2cm}
\noindent \textit{Building Resilience as a Sociotechnical Property.}\\
Resilience is more than the ability to recover from failure. A resilient agentic AI system must integrate prevention, preparation, detection, recovery, and adaptation, each of which becomes more challenging in a setting where failures may not yet have been seen. Existing harm taxonomies often do not %fully capture
anticipate the kinds of damage agentic AI systems can produce, and learning from each new failure introduces its own risks. An AI agent that adjusts its behavior after a mistake may create new vulnerabilities in the process or alter its behavior in unintended or undesirable ways.
%Resilience is more than the ability to recover from a failure. A resilient agentic system must integrate prevention, preparation, detection, recovery, and adaptation, each of which presents distinct challenges in a setting where failures may be novel, undocumented, or invisible until they cause harm. Prevention requires anticipating failure modes that have not yet occurred; preparation requires standing infrastructure capable of absorbing those failures when they do; detection requires the legibility and auditing capabilities discussed above; recovery requires that the system return to a safe state without losing the user's accumulated context; and adaptation requires that lessons from each failure improve future behavior without introducing new vulnerabilities through the learning process itself.

Resilience cannot be a purely technical property. The humans involved are not only end users. Developers, operators, regulators, and others affected by the actions of an AI agent all have a role in detecting failures, deciding what counts as safe, and shaping what the AI agent learns over time. Placing humans out of the resilience loop in the name of efficiency risks producing agentic AI systems that drift from their original purpose. At the same time, involving humans in a na\"{i}ve manner risks reproducing the consent-fatigue problems described earlier (see \S\ref{sec:theme2}).

Finally, the resource requirements for resilience are often overlooked in security and privacy discussions. The computational and environmental costs of running, monitoring, and auditing agentic AI systems are substantial, and questions about hardware coordination, energy use, %  across heterogeneous devices, the energy footprint of continuous inference, and the 
and sustainability of fully agent-mediated infrastructures intersect with security and privacy in ways that are not yet well theorized. A resilient agentic AI ecosystem must be one that can be sustained over time, not only against adversaries but also against the material constraints of the 
%systems
environment in which it operates.
%ISabel
%Addition: Resilience should be understood not as the elimination of failure but as the capacity of an agentic ecosystem to continue operating safely despite inevitable failures while learning from them without compromising security, privacy, or accountability.

\section{Conclusion}

This article set out the most pressing security and privacy challenges that arise as AI moves from passive assistance toward proactive, autonomous action. The challenges are organized around four themes: accountability, governance and liability; consent, data, and contextual integrity; human oversight, personalization, and vulnerability; and transparency, resilience, and failure recovery. A common pattern runs through them. The foundational concepts of security and privacy research---consent, contextual integrity, traceability, certification, vulnerability---were developed for bounded, deterministic, attributable systems. Agentic AI violates these foundational assumptions and, therefore, requires reconceptualisation rather than refinement. Whether the existing vocabulary of security and privacy is sufficient for the agentic AI era or whether new categories of risk need to be named and theorized alongside existing ones remains an open question for the research community to address.

% conference papers do not normally have an appendix

% use section* for acknowledgment
\ifCLASSOPTIONcompsoc
  % The Computer Society usually uses the plural form
  \section*{Acknowledgments}
\else
  % regular IEEE prefers the singular form
  \section*{Acknowledgment}
\fi

We thank David Rodriguez, Jaime Andr\'{e}s Rinc\'{o}n Arango, Jose M. Del Alamo, Juan Caballero, Nadin Kokciyan, Luis B\'{u}rdalo, Pinar Yolum, Simone Fischer-H\"{u}bner, Verena Distler, and Yixin Zou for participating in the horizon scanning activity. % and for their valuable contributions to the discussions. 
The horizon scanning activity was funded by the INCIBE's strategic SPRINT (Seguridad y Privacidad en Sistemas con Inteligencia Artificial) C063/23 project with funds from the EU-NextGenerationEU through the Spanish government's Plan de Recuperación, Transformación y Resiliencia. The authors used a locally deployed Qwen3.5-27B~\cite{qwen35blog} large language model to assist with summarizing moderator notes and clustering themes with semantically similar ones. %In both cases, several authors went through the notes and the clusters produced to validate that the summary accurately reflected the notes and nothing was left out, and that the clusters produced made sense semantically. %No data were submitted to external AI services.

% trigger a \newpage just before the given reference
% number - used to balance the columns on the last page
% adjust value as needed - may need to be readjusted if
% the document is modified later
%\IEEEtriggeratref{8}
% The "triggered" command can be changed if desired:
%\IEEEtriggercmd{\enlargethispage{-5in}}

% references section

% can use a bibliography generated by BibTeX as a .bbl file
% BibTeX documentation can be easily obtained at:
% http://mirror.ctan.org/biblio/bibtex/contrib/doc/
% The IEEEtran BibTeX style support page is at:
% http://www.michaelshell.org/tex/ieeetran/bibtex/
%\bibliographystyle{IEEEtran}
% argument is your BibTeX string definitions and bibliography database(s)
%\bibliography{IEEEabrv,../bib/paper}
%
% <OR> manually copy in the resultant .bbl file
% set second argument of \begin to the number of references
% (used to reserve space for the reference number labels box)
% \begin{thebibliography}{1}

% \bibitem{IEEEhowto:kopka}
% H.~Kopka and P.~W. Daly, \emph{A Guide to \LaTeX}, 3rd~ed.\hskip 1em plus
%   0.5em minus 0.4em\relax Harlow, England: Addison-Wesley, 1999.

% \end{thebibliography}

\bibliographystyle{IEEEtran}
\bibliography{ref}

% that's all folks
\end{document}